# Lifetime Study of COTS ADC for SBND LAr TPC Readout Electronics

H.Chen, J.Evans, J.Fried, S.Gao*, F, Liu, G. Miller, K. Mistry, J. Pater, A.Szelc, V.Radeka, J.Zhang

*Abstract*—Short Baseline Near Detector (SBND), which is a 260-ton LAr TPC as near detector in Short Baseline Neutrino (SBN) program, consists of 11,264 TPC readout channels. As an enabling technology for noble liquid detectors in neutrino experiments, cold electronics developed for extremely low temperature (77K–89K) decouples the electrode and cryostat design from the readout design. With front-end electronics integrated with detector electrodes, the noise is independent of the fiducial volume and about half as with electronics at room temperature. Digitization and signal multiplexing to high speed serial links inside cryostat result in large reduction in the quantity of cables (less outgassing) and the number of feed-throughs, therefore minimize the penetration and simplify the cryostat design. Being considered as an option for the TPC readout, several Commercial-Off-The-Shelf (COTS) ADC chips have been identified as good candidates for operation in cryogenic temperature after initial screening test. Because Hot Carrier Effects (HCE) degrades CMOS device lifetime, one candidate, ADI AD7274 fabricated in TSMC 350nm CMOS technology, of which lifetime at cryogenic temperature is studied. The lifetime study includes two phases, the exploratory phase and the validation phase. This paper describes the test method, test setup, observations in the exploratory phase and the validation phase. Based on the current test data, the preliminary lifetime projection of AD7274 is about $6.1 \times 10^6$ years at 2.5V operation at cryogenic temperature, which means the HCE degradation is negligible during the SBND service life.

## I. INTRODUCTION

THE Short Baseline Near Detector (SBND) is one of three liquid argon (LAr) neutrino detectors sitting in the Booster Neutrino Beam (BNB) at Fermilab as part of the Short Baseline Neutrino (SBN) program. The detector is in a cryostat holding 260-ton LAr and consists of 4 full-size Anode Plane Assembles (APAs) plus 2 Cathode Plane Assemblies (CPAs), as shown in Fig. 1, which gets 2 × 2m drift regions and 11, 264 TPC (Time Projection Chamber) readout channels [1]. As an enabling technology, cold electronics developed for cryogenic temperature (77K – 89K) operation makes possible an optimum balance among various design and performance requirements for such large sized detectors [2]. There are two main advantages of cold electronics. First, large detectors used for neutrino experiment requires very low noise performance to meet their physical goals. Cold electronics decouples the electrode and cryostat design from the readout design. With electronics integral with detector electrodes, the input capacitance of signal cable is negligible, which results in the noise independent of the fiducial volume (signal cable lengths). Meanwhile, benefit from charge carrier mobility in silicon increasing and thermal fluctuations decreasing with kT/e, the noise of CMOS front end (FE) ASIC significantly decreases at cold temperature. Second, signal digitization and multiplexing to high speed links inside the cryostat result in large reduction in the quantity of cables (less outgassing) and the number of feed-through penetrations, also giving the designers of both the TPC and the cryostat the freedom to optimize the detector configurations. Therefore, SBND chooses cold electronics as LAr TPC readout solution. As shown in Fig. 2, CMOS cold FE ASIC is placed close to the wire electrodes to detect charge signal, followed by the ADC for signal digitization. Digitized signal will be further organized and transmitted to the downstream readout and Data Acquisition (DAQ) system [3].

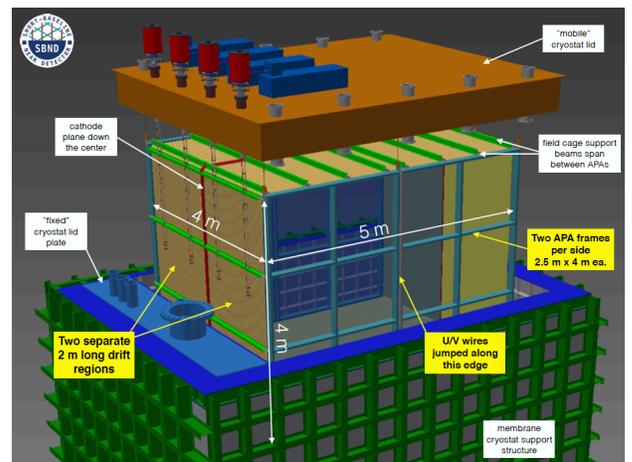

Fig. 1. Overview of SBND LAr TPC. 2 APAs are on either side; the sensing area of each APA is 2.5m x 4 m. With Two CPAs in central, two separate 2m long drift regions are formed.

There nearly wasn't any commercial semiconductor device specifically designed for cryogenic temperature operation. After studies of the transistor characteristics and lifetime at

This work was supported in part by the U.S. Department of Energy.
H.Chen, J. Fried, S. Gao, V.Radeka, J. Zhang are with Brookhaven National Laboratory, NY, 11973, USA.
F.Liu is with the Key Laboratory of Particle & Radiation Imaging, Ministry of Education, Beijing, China, and the Department of Engineering Physics, Tsinghua University, Beijing, 10084 China.
J.Evans, G. Miller, K. Mistry, J. Pater, A.Szelc are with University of Manchester, Manchester, M13 9PL, United Kingdom.
Corresponding author: S. Gao (email: sgao@bnl.gov )

cryogenic temperature, BNL designed an FE ASIC suitable for 77K–300K operation and long lifetime with lower power consumption [4]. Meanwhile, various commercial CMOS devices have been screened to find survivors at cryogenic temperature. Two CMOS devices, a low voltage regulator from TI and an Altera Cyclone IV FPGA are qualified for cold electronics development. Since high-resolution and low power consumption SAR architecture ADC becomes widely available, we started to characterize the performance of Commercial-Off-The-Shelf (COTS) ADC at cryogenic temperature since June 2017. Several COTS ADC chips have been identified as good candidates for operation at cryogenic temperature after initial screening test [5]. One candidate, ADI AD7274 fabricated in TSMC 350nm CMOS technology, of which lifetime at cryogenic temperature has been studied since September 2017. The following sections describe the test method, test setup and test observations, and finally give the lifetime projection.

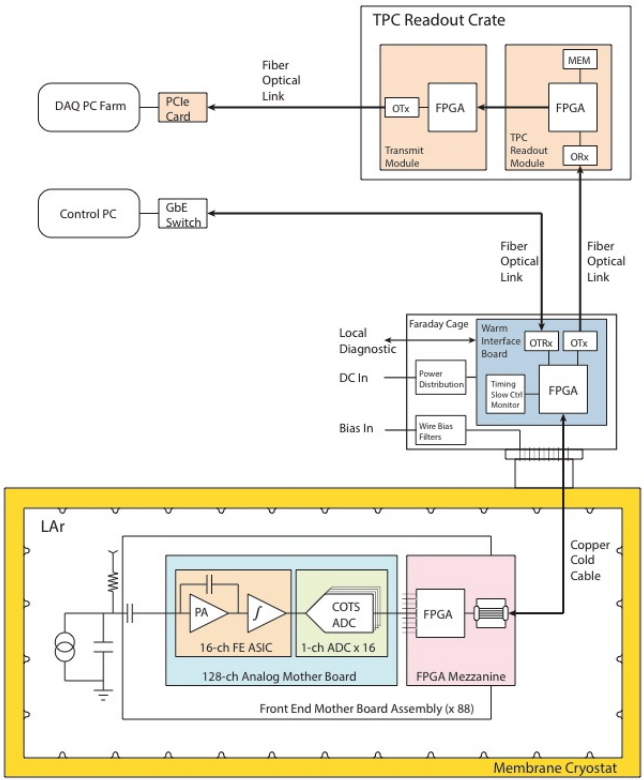

Fig. 2. Diagram of SBND TPC readout electronics. 11, 264 TPC channels are read out by 88 front end mother board assemblies (FEMBs). Each FEMB consists of an analog motherboard (AM) and a FPGA Mezzanine (FM). AM has 8× 16-ch FE ASICs to pre-amplify and shape detector signals, 128 single channel ADC to digitize signals. FM multiplexes and transmits digitized signals over 4× 1Gb/s links to warm interface electronics and then to DAQ system.

## II. CMOS Lifetime At Cryogenic Temperature

Past studies [6] [7] indicate that most of the major failure mechanisms are strongly temperature dependent and become negligible at cryogenic temperature, such as electro-migration, stress migration, time-dependent dielectric breakdown and negative-bias temperature instability. The only remaining mechanism that may affect the lifetime of CMOS devices at cold is the degradation (aging) due to channel Hot Carrier Effects (HCE). Lifetime due to HCE aging is a limit defined by a chosen level of monotonic degradation in e.g., drain current, transconductance, and threshold voltages. The aging mechanism does not result in sudden device failure. The device "fails" if a chosen parameter gets out of the specified circuit design range. Two different measurements can be used to evaluate the HCE aging: accelerated lifetime measurement under severe electric field stress by the drain-source voltage ($V_{ds}$), and a separate measurement of the substrate current ($I_{sub}$) as a function of $1/V_{ds}$. The former verifies the canonical very steep slope of the inverse relation between the lifetime and the substrate current, $\tau \propto I_{sub}^{-3}$ and the latter confirms that below a certain value of $V_{ds}$ a lifetime margin of several orders of magnitude can be achieved for the cold electronics in LAr TPC readout. The front end ASIC designed for SBND and DUNE falls naturally into this domain, where hot-electron effects are negligible.

## III. COTS ADC Lifetime Study Method

The CMOS lifetime study at 300K and 77K has been carried out in the R&D of cold electronics for DUNE experiment since 2008, an IEEE paper was published in 2013 [6]. The paper shows that TSMC 180nm node projects a lifetime in excess of $10^3$ years at 77K and nominal $V_{ds}$ = 1.8V under DC conditions. We would expect the same, or better for TSMC 350nm used by AD7274, based on the lower peak electric field in the channel (and lower hot-carrier effect). There is no internal regulator inside AD7274 so that the accelerated lifetime measurement under severe electric field stress by the drain-source voltage ($V_{ds}$) can be devised.

The lifetime study of COTS ADC takes place in two different phases, the exploratory phase and the validation phase. The most time and effort focus on the exploratory phase, we would gain (or lose) confidence based on the test results in this phase. During the exploratory operation phase, fresh ADC samples will be stressed with higher than nominal voltage, such as 5.5V, while power consumption (current drawn) will be monitored continuously. Periodically the ADC sample will be operated at nominal voltage for performance characterization test, where both DNL (Differential Non-Linearity) and INL (Integral Non-Linearity) will be monitored and analyzed. The development of lifetime criteria will be an iterative process, tailored to ADC technology, based on the test data to be collected and analyzed. In the validation phase, more devices will be tested following the criteria developed in the exploratory phase. The goal is to collect more data to validate what we would have learned in the exploratory phase, and further to conclude the lifetime study.

## IV. COTS ADC Lifetime Test Setup

A COTS ADC lifetime study test setup during the stress test in $LN_2$ is shown in Fig. 3. A relay board is realized to quickly switch power between stress mode and performance characterization mode. During the stress mode, power of ADC will be supplied directly from the Source Measure Unit (SMU) that can provide precision current monitoring to µA level. During the performance characterization mode, low noise regulator ADM7151 evaluation boards are used to provide low noise power to ADC. Only one fresh sample is being stress tested on the COTS ADC test board, which is being read out through an FPGA mezzanine. Both COTS ADC test board and FPGA mezzanine are submerged and operating in a $LN_2$ dewar. The data is sent out from FPGA mezzanine to Arria V evaluation board through a MiniSAS cable, from there a Gigabit Ethernet cable is connected to PC for data acquisition through UDP protocol. Python scripts have been prepared for automatic data taking and data analysis. No intervention for test setup is required during the test except $LN_2$ filling in dewar.

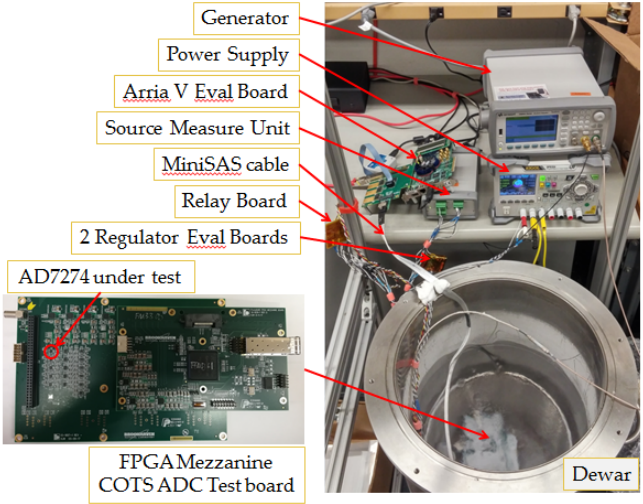

Fig. 3. COTS ADC lifetime test stand. The test stand is used in the exploratory phase. The ADC test board was slightly upgraded to support up to 6.0V stress voltage for the validation phase.

Three similar AD7274 lifetime study test setups have been built. For the current monitoring, ADC is running at 2Msps with full-scale triangle input in $LN_2$. During the stress test, both $V_{DD}$ and $V_{REF}$ of ADC are applied with same voltage; during the nominal operation for performance characterization, $V_{DD}$ = 2.5V, $V_{REF}$ = 1.8V.

## V. OBSERVATION IN THE STRESS TEST

The lifetime, which dues to HCE aging at both the cryogenic and room temperatures, is a limit characterized by a predictable and a very gradual and monotonic degradation mechanism, can be defined by any arbitrary but consistent criteria. The most efforts spend in the exploratory phase is to find the monotonic degradation criteria [8].

The DNL of ADC sample under test has been characterized. The sigma of ADC DNL distribution at 2Msps along the stress test is shown in Fig. 4. A sample tested for >800 hours with nominal operation voltages ($V_{DD}$ = 2.5V and $V_{REF}$ = 1.8V) doesn't show any significant performance change, which confirms the observation of DNL sigma change is caused by stress test for samples under the stress voltages. The results of DNL performance of ADC samples under stress is promising, overall, the change of sigma of DNL distribution is less than 0.2 LSB through the stress test, and none of ADC samples failed, which means the HCE degradation has little contribution to ADC performance. However non-monotonic behavior of sigma of DNL is observed and re-producible. DNL performance variation through the cold stress test is different from samples with same stress voltages, as well as samples with different stress voltages. So sigma of DNL is not suitable for the lifetime projection.

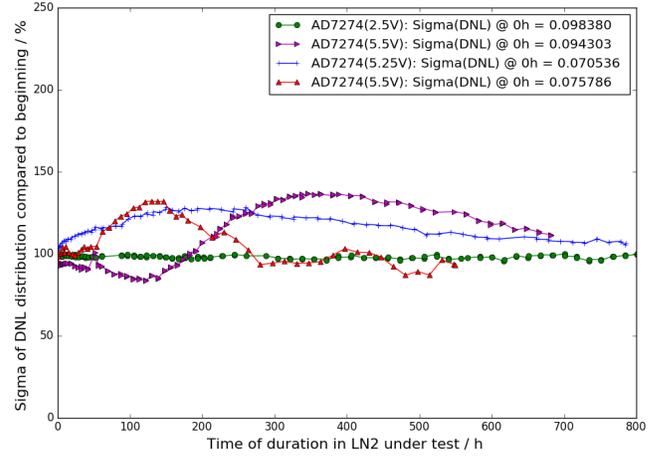

Fig. 4. Sigma of DNL distribution along the cold stress tests. The sample with nominal voltage (2.5V) doesn't show significant performance change. Non-monotonic behavior of sigma of DNL is observed from samples with stress voltages.

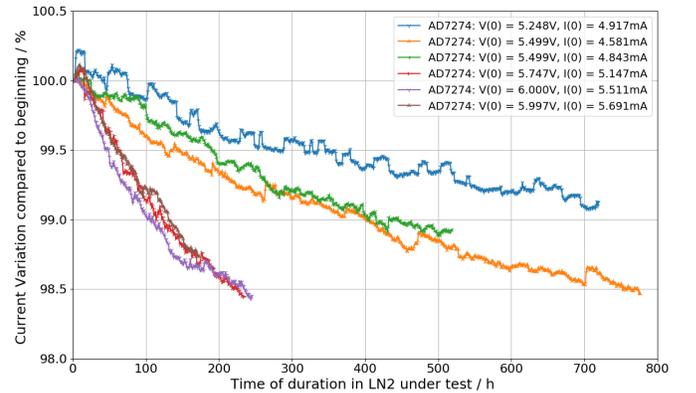

Fig. 5. Current of VDD vs. Time of duration in $LN_2$. The current degradation is faster with higher stress voltages.

Meanwhile, the current monitoring of COTS ADC in the stress test shows the expected results. In order to see the variation of current, a high precision current measurement method is required, as well as careful designed power distribution scheme. A SMU with µA level current monitoring is applied, accompanying with triangle waveform as input to ADC to mitigate current fluctuation issue caused by the reference voltage of ADC ($V_{REF}$). As shown in Fig. 5, current of $V_{DD}$ decreases consistently with different stress voltages. The current decreases faster with higher stress voltages, regardless of the test stand. Two samples stressed at 5.5V, tested by BNL and Manchester University independently, overlap quite well.

## VI. COTS ADC Lifetime Projection

Due to some setup issues, such as computer crash, data loss, SMU or power supply reboot, broken connection, lack of $LN_2$, etc. along the development of COTS ADC lifetime test setup, 6 ADC samples with continuous current monitoring data at cryogenic temperature are used to project the lifetime of ADC. 1% of $I_{VDD}$ (current of $V_{DD}$) drop is determined as the degradation criteria. The sample stressed at 5.25V and stopped after ~710 hours is estimated to have 1% current drop at 800 hours. Based on the empirical equation, as in (1), the preliminary ADC lifetime projection is plotted in Fig. 6

$$\log_{10} \tau \propto 1/V_{ds} \qquad (1)$$

The lifetime of AD7274 at the technology node with 3.6V is projected to ~160 years. Reduced $V_{DS}$ to 2.5V results in a very long extrapolated lifetime to ~6.1 × $10^6$ years, which means HCE degradation is negligible. As a principle, the cold electronics for LAr TPCs should be designed for a lifetime one or more orders of magnitude longer than the required service life (e.g., > 100 years for SBND or >300 years for DUNE). The lifetime study shows the AD7274 is cold qualified devices to remain outside of the region of HCE degradation.

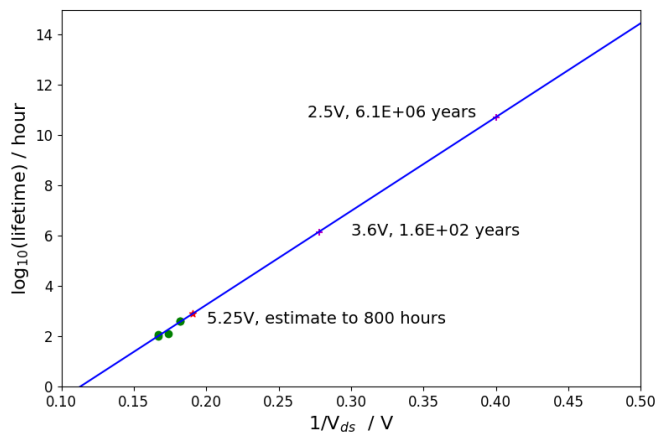

Fig. 6. A preliminary ADC lifetime projection. Validation phase of the stress test is still ongoing to collect more data to check the margin of the ADC design and examine the difference from different batches, and get a sense of impact of process variation.

## VII. Conclusion

Cold electronics makes possible an optimum balance among various design and performance requirements for SBND LAr TPC detector. The COTS ADC cold lifetime study presents a very promising result. Preliminary lifetime projection of AD7274 is ~6.1 × $10^6$ years at 2.5V operation. The HCE will be negligible for the ADC used in SBND, and we'll be stay out of HCE during the detector service life. Though validation phase of AD7274 lifetime study is still ongoing with the aim to publish this study, SBND collaboration has made the decision to use AD7274 as the cold ADC option in the front end readout electronics. The COTS ADC work also benefits the future program, which serves as a potential backup for DUNE far detector, cold qualification techniques are useful for future dark matter and neutrino experiments.